\documentclass[%
reprint,
superscriptaddress,
 amsmath,amssymb,
 prl,
showkeys,
nofootinbib
]{revtex4-2}

\usepackage{graphicx}% Include figure files
\usepackage{dcolumn}% Align table columns on decimal point
\usepackage{bm}% bold math
\usepackage{hyperref}% add hypertext capabilities
\usepackage{color} %  for colored comments
\begin{document}

%\preprint{APS/123-QED}

\title{
Energy distribution of inelastic gas in a box is dominated by a power law  -- a derivation in the framework of sample space reducing processes
}

\author{Stefan Thurner}
\affiliation{Section  for  the  Science  of  Complex  Systems,  CeMSIIS, Medical  University  of  Vienna,  Spitalgasse  23,  A-1090,  Vienna,  Austria
}
\affiliation{Complexity  Science  Hub,  Vienna  Josefst\"{a}dterstrasse  39,  A-1090  Vienna,  Austria}
\affiliation{Santa  Fe  Institute,  1399  Hyde  Park  Road,  Santa  Fe,  NM  87501,  USA}

\author{Jan Korbel}
\affiliation{Section  for  the  Science  of  Complex  Systems,  CeMSIIS, Medical  University  of  Vienna,  Spitalgasse  23,  A-1090,  Vienna,  Austria
}
\affiliation{Complexity  Science  Hub,  Vienna  Josefst\"{a}dterstrasse  39,  A-1090  Vienna,  Austria}

\author{Rudolf Hanel}
\affiliation{Section  for  the  Science  of  Complex  Systems,  CeMSIIS, Medical  University  of  Vienna,  Spitalgasse  23,  A-1090,  Vienna,  Austria
}
\affiliation{Complexity  Science  Hub,  Vienna  Josefst\"{a}dterstrasse  39,  A-1090  Vienna,  Austria}

\date{\today}% It is always \today, today,

\begin{abstract}
We use the framework of sample space reducing processes (SSR) as an alternative to Boltzmann equation based approaches, to derive the energy and velocity distribution functions of an inelastic gas in a box as an example for a dissipative, driven system. SSR processes do not assume molecular chaos and are characterized by a specific type of eigenvalue equation whose solutions represent stationary distribution functions. The equations incorporate the geometry of inelastic collisions and a driving mechanism in a transparent way. Energy is injected by boosting particles that hit the walls of the container to high energies. The numerical solution of the resulting equations yields power laws over the entire energy region. The exponents decrease with the driving rate from about $2$ to below $1.5$ and depend on the coefficient of restitution. Results are confirmed with a molecular dynamics simulation in 3D with the same driving mechanism.  We believe that these distribution functions are observable in experimental situations; the abundance of power laws in driven dissipative systems might be a sign of the generality of the result.
\end{abstract}

%Use showkeys class option if keyword display desired
\maketitle

%\section{Introduction}

Driven dissipative systems remain a challenge for statistical physics since well more than a century. Even in their simplest form, such as an inelastic gas in a box with a simple driving mechanism that re-introduces dissipated energy during wall collisions, they have not been solved for stationary conditions. The equivalent of the Maxwell-Boltzmann distribution for elastic gases %in a fixed volume at a given temperature 
is still not fully known for inelastic gases. Much less is known for  dissipative systems that abound in nature, including examples as diverse as non-equilibrium thermodynamics \cite{Lebon2008}, granular matter, turbulent flow \cite{Munson1990}, self-organization \cite{Schieve1982}, the earth \cite{Kleidon2012}, and living systems \cite{Goldbeter2018}.  

What has been understood in inelastic gases for several decades, however, is that due to the inelasticity in the collisions, generally energy and velocity distributions are non-Maxwellian \cite{Kudrolli1997,Noije2003,Rouyer2000,Barrat2002}. Since the 1970s it was noted in many contributions in a wide range of fields that power laws play an important role. Power law solutions in the Boltzmann equation were found in numerous contributions \cite{Kats1976,Baldassarri2002,Huang2019}. 
Understanding the scaling velocity distributions in inelastic particle systems was pushed in the understanding of non-linear Boltzmann equations and the presence of multiscaling \cite{Ben-Naim2000,Ben-Naim2002,Ben-Naim2005}. In the latter power laws are derived analytically with an exponent, $\gamma(c_r,D)$ that depends on the restitution  coefficient, $c_r$, and the spatial dimension, $D$. It can be derived again from the Boltzmann equation. However, as we understand better now in this work, the power law found in \cite{Ben-Naim2005} applies to the extremely energetic particles only, but not to the entire energy region. 
The reason for this is the assumption of a weak version of molecular chaos, i.e. that post-collision velocities are not correlated when one particle has extremely high and the other low energy. We will see that the high energy tail of the velocity distribution can be obtained without the molecular chaos assumption by an alternative approach to inelastic gases.

Here we suggest to use an entirely different approach to inelastic gases that is not based on the (non-linear) Boltzmann equation but on sample space reducing (SSR) process \cite{Corominas-Murtra2015}. The SSR framework has been shown useful to deal with processes that violate detailed balance. We compute the energy (and velocity) distribution functions over the entire energy/velocity region for an inelastic gas in a box coupled to a simple driving mechanism, where energy is injected through those particles that hit the walls of the containing box. We understand the effects of the driving rate and check the analytical results with molecular dynamics (MD) simulations.  

Driven systems are typically composed of a driving and a relaxation part, often in arbitrarily complicated ways. When systems relax toward lower (energy) states, this usually happens as a sample space reducing (SSR) process. The corresponding distribution functions are relatively easy to compute, once the details of the driving process are specified \cite{Corominas-Murtra2018}. For simple driving processes, SSR processes were found to exhibit universal power law statistics of visiting frequencies of the systems' states, regardless of the details in the relaxation dynamics.

%\subsection{Dissipative systems and the SSR argument}

Dissipating processes such as inelastic collisions in a box are sample space reducing (SSR) processes in the following sense. Without driving, systems relax towards lower (energy) states over time. Assume that a system has $M$ states that can be ordered or ranked (such as energy), labelled by $i\in {1,2,\cdots, M}$. The probability distributions of finding the system in (energy) state $i$ are given by the eigenvalue equations of the following type, 
\begin{equation} 
 p(i) = \sum_j p(i | j) p(j) \, , 
 \label{eq:ev}
\end{equation}
where $p(i | j)$ is the transition probability that the system passes from state $j$ to a lower state $i$. In the simplest case, 
\begin{equation}
p(i|j)=\left\{
\begin{array}{cl}
	\frac{q_i}{ \sum_{k < j} q_k}& {\rm for}\; i<j \\
	0 			& {\rm for}\; i \geq j \, , \nonumber
\end{array}
\right.
\label{eq:P(a_k|a_i)}
\end{equation}
where the system jumps to any lower state with the weight, $q_i$. It defines the probability for visiting state $i$. In the simplest case, whenever the lowest state is reached, the system is restarted at any randomly chosen energy level (driving process). The solution to Eq. (\ref{eq:ev}), the distribution of visiting frequency, is an exact power law with exponent $-1$, sometimes referred to as Zipf's law, $p(i) \sim i^{-1}$, see \cite{Corominas-Murtra2015}. If the system is restarted before it reaches the ground state, say with probability $1-\lambda$ (with $0\le\lambda\le1$) at every timestep, the resulting distribution remains an exact power, however, now with exponent, $-\lambda$. Remarkably, this is true for a huge class of choices of $q_i$, the result is always an exact power law \cite{Corominas-Murtra2016}. Processes of this type are called sample space reducing processes since for the majority of the transitions 
the number of possible reachable states (sample space) shrinks as the process unfolds. 

Elastic collision processes (with energy conservation) can be described as SRR processes. For example, imagine a high-velocity particle with initial kinetic energy, $E_0$, crashing into a box of resting classical particles all of the same mass that are sparsely distributed.  When following the initial particle, after the first collision with a resting particle, it goes to a lower kinetic energy, $E_1<E_0$. The formerly resting particle now has kinetic energy and can kick other resting particles. For simplicity, we assume that it will never kick the initial particle again.  The initial particle will lose energy along a sequence of $n$ collisions, and we have a sample space reducing process, $E_n<....<E_1<E_0$. After some time, the initial particle will leave the box (no boundary). The system is driven by shooting particles with $E_0$ into the box. The energy distribution of the particles can be computed analytically by solving the eigenvalue equation, which again yields an exact power law with exponent $-2$, see \cite{Corominas-Murtra2017}. Here we show that the framework based on SSR processes  allows us to also treat ensembles of inelastic collisions, in particular, the equivalent to the Maxwell-Boltzmann distribution for inelastic gases can be computed. 

\begin{figure}[t]
\hspace{-4.5cm} { \small {\textsf{ (a)} }} \\
\includegraphics[scale=0.08]{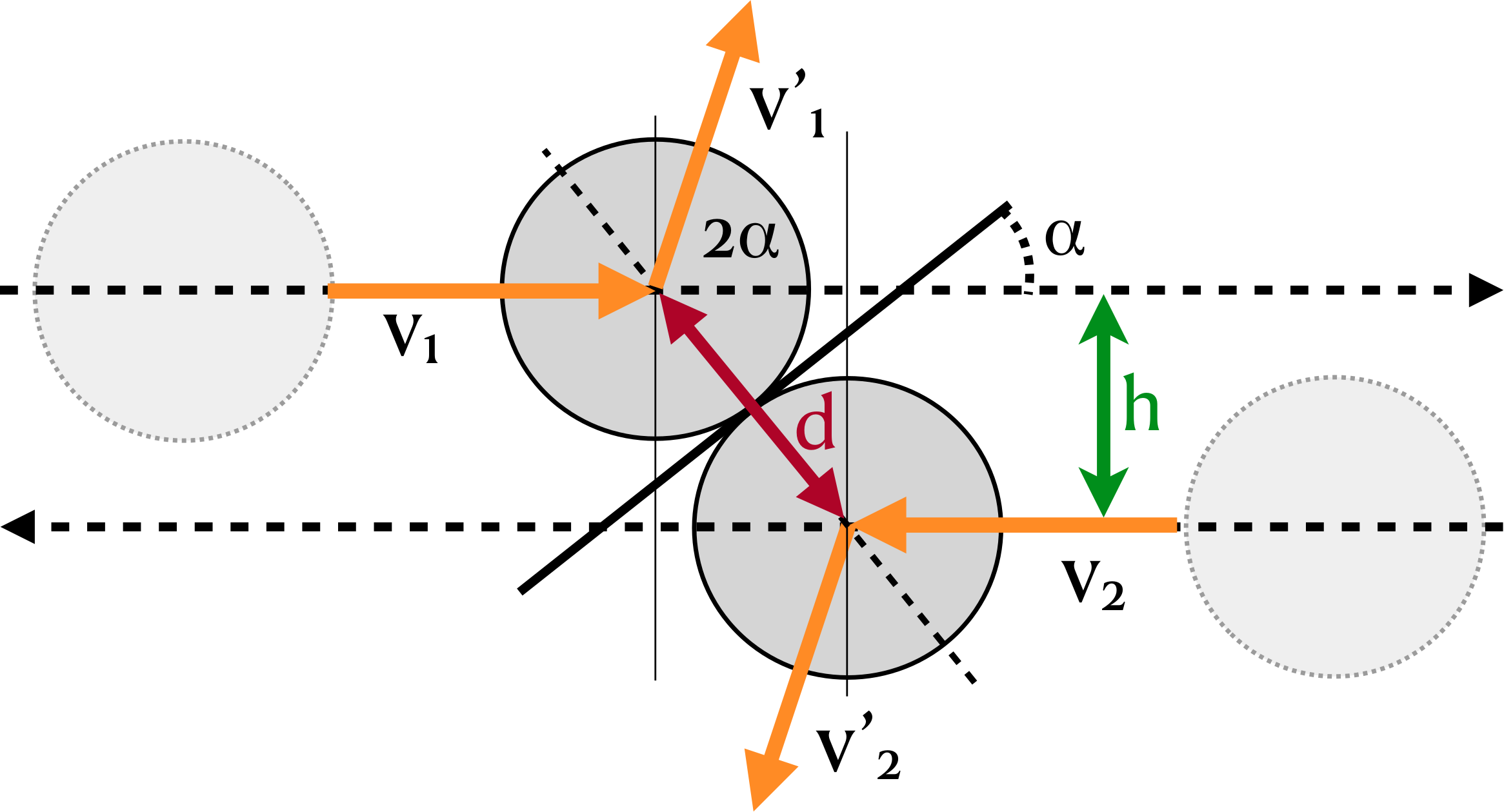}
\includegraphics[scale=0.12]{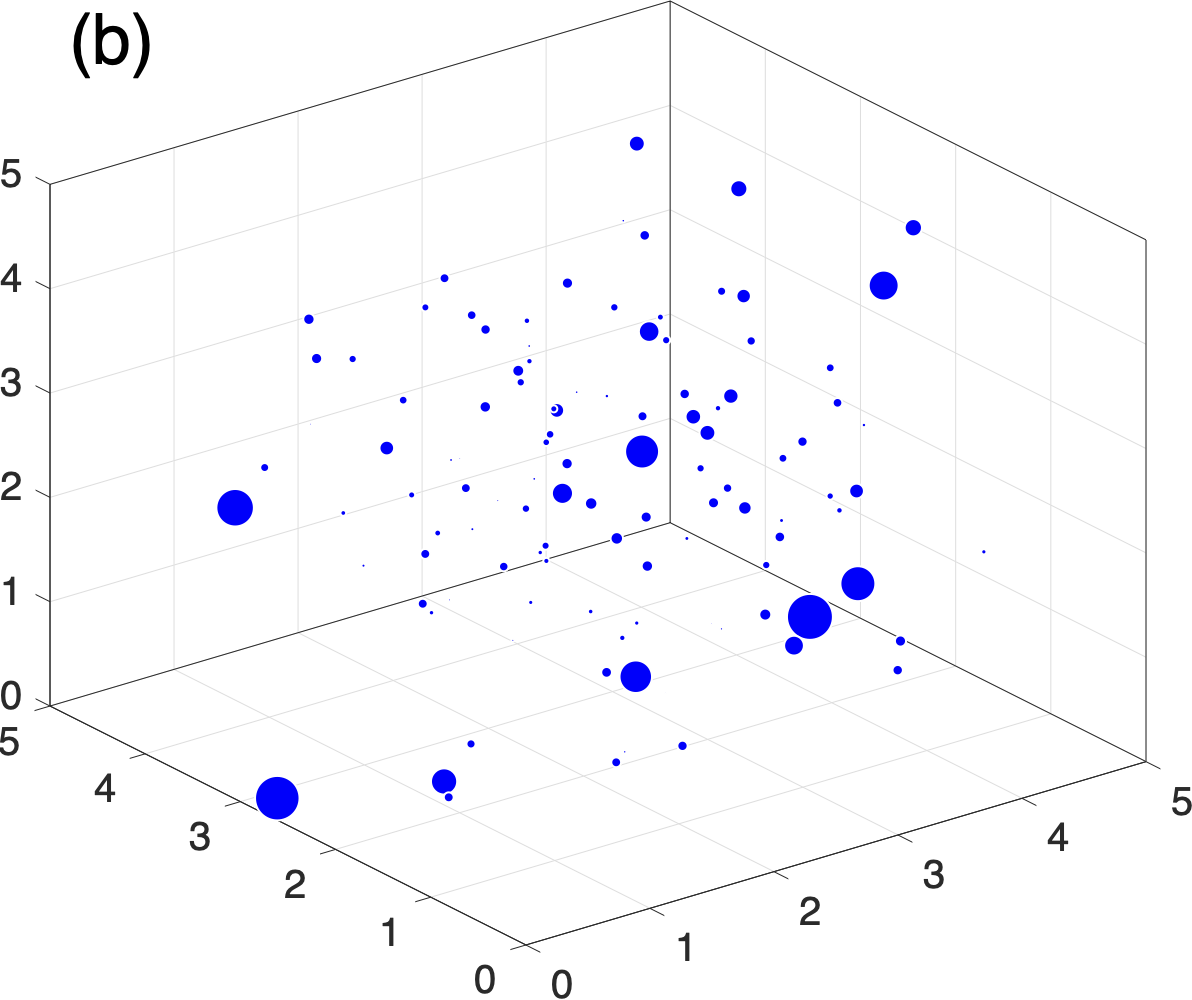}
\caption{
(a) Notation for the inelastic collision in the center of mass frame. 
(b) Particles collide inelastically with each other in a box, walls reflect elastically. 
At the wall-collisions energy gets reintroduced with probability $\eta$ to a fixed energy, $E_{\rm charge}=5$.
The plot shows $N=125$ particles after 10,000 collisions, their size represent their kinetic energy, 
$d=0.6$, $c_r=0.7$. Particles are not uniformly distributed within the box, slow ones lump together in a cluster. 
}
\label{fig:1}
\end{figure}

%\subsection{Inelastic particles in a box}

We consider $N$ identical classical particles with diameter, $d$, and unit mass, $m=1$, in a three dimensional box of size $L$. Particles collide with each other inelastically with a coefficient of restitution, $c_r$. For the geometry of the collision, see Fig. \ref{fig:1} (a). In the center of mass frame, two particles, $1$ and $2$, with incoming velocities $v_1$ and $v_2$ collide at an angle $\alpha$. In this frame, $ v^{cm} = (  v_1 + v_2 )/2$, the relative distance vector is
$\hat r = ( x_2 - x_1)/|  x_2 - x_1 |$ and the velocities after the collision are 
\begin{align} 
 v_1' &= ( v_1  -  [( v_1- v_2) \cdot \hat r]  \hat r-   v_{cm}) c_r+ v_{cm} \nonumber \\
 v_2' &= ( v_2 +  [( v_1- v_2) \cdot \hat r]  \hat r-   v_{cm}) c_r+ v_{cm}  \, .
\end{align}
$c_r=1$ means elastic collision, for $0<c_r<1$, kinetic energy is no-longer conserved, $E'<E^{\rm before\, coll}$. Particles are reflected elastically at the walls of the box. The system dissipates energy in every particle-particle collision (except for exactly tangential hits). 
%If no energy is re-introduced the system goes to lower and lower energies.  

There are many ways to re-introduce the dissipated energy to arrive at a stationary situation. In the spirit of an energy bath, the driving process could be realized such that particles that hit the wall are boosted to a high energy level,drawn from the driving distribution $\rho_{\rm charge}(E)$ (wall has a temperature and transmits it to particles when in contact). Alternatively, randomly chosen particles could be injected with an energy from the same distribution (for example, by shining laser pulses into the gas of particles).   
Many other possibilities can be imagined and implemented.

For the following analytical computations we chose a driving scheme, where particles whenever they hit a wall, with a 
probability $\eta$ are set to a fixed kinetic energy, $\rho_{\rm charge}(E)=\delta(E-E_{\rm charge})$; the direction of the particle left unchanged (up to reflection). In terms of velocity, a charging process for particle $1$ means
$ v_1 = |{v}_1| {\bf{v}}_1   \to  v'_1  =   ( 2 E_{charge} /m   )^{\frac12} {\bf{v}}_1$,
where $\bf{v}$ is the unit velocity vector. The details of the driving process are known to be relevant for the resulting energy distribution functions, especially the driving rate plays a crucial role \cite{Corominas-Murtra2018}. We define the driving rate, $r$, as the ratio of energy re-charging events per particle-particle collision. Note that $r$ depends not only on $\eta$ but also on the geometry of the system, 
in particular, the particle diameter and the particle density in the box.

Figure \ref{fig:1} (b) shows a snapshot of an inelastic gas in a box. The size of the particles represents their kinetic energy. There appears a cluster of low-energy particles at the lower right corner in the back of the box. Particles with high energy have a higher chance of getting re-charged in a wall collision.

%\section{Model}

The idea is to compute the energy distribution, $\rho(E)$, of a driven inelastic gas in a stationary state by solving an eigenvalue equation of the type given in Eq. (\ref{eq:ev}), in particular
\begin{equation}
\rho(E')=\int_{E_0}^{E^*}dE\rho(E'|E,c_r)\left((1-\xi)\rho(E)+ \xi \rho_{\rm charge}(E)\right)\, ,
\label{eq:e}
\end{equation}
for a specific geometry of inelastic collisions. $\rho(E)$ is the stationary energy distribution function and $\rho_{\rm charge}(E)$ is the energy distribution of particles just after a driving event. The internal energy of the system is $U = (1- \xi) \langle E \rangle _{\rm post} + \xi \langle E \rangle _{\rm charge}$, where $\langle E \rangle_{\rm post}$ and $\langle E \rangle_{\rm charge}$ denote the expectation values for the energy distribution and the driving energy distribution (energy source or bath). Per unit time, a fraction of $\xi$ particles are drawn from $\rho(E)$ and are replaced with a new energy, drawn from $\rho_{\rm charge}(E)$. The other fraction of particles, $1-\xi$, undergo particle-particle collisions and receive no energy charge from the source. Since we measure the driving rate, $r$, as the number of driving-kicks per particle-particle collision, within a time span $\tau$, where each of the $N$ particles collide once on average, we get $N/2$ particle-particle collisions, and $r = 2\xi$. $\tau$ is the average inter-particle collision time that we assume to be independent of the particular particle energy $E$. For Eq. (\ref{eq:e}) to make sense, we assume that every particle, irrespective of its energy or position, is driven with identical probabilities. To compute it, we need the single-particle energy transition probabilities that depend on the geometry of the collision. For the details and the derivation, see SI Text 1.

The single-particle transition probability in 3D, is obtained by integrating over the involved variables,  $\zeta$, $\alpha$, and $\phi$ and their respective probability functions, $g(\zeta)=\sin(\zeta)/2$,  $f(\alpha)=|\sin 2\alpha |$, and $r(\phi)={1}/{\pi}$, (definitions, see SI Text 1)
\begin{align}
&\rho  (E_1'|E_1,c_r) = \frac{1}{Z(E_1)} \int_{0}^\pi d\zeta g(\zeta) \int_0^{\pi} d\alpha f(\alpha) \int_0^{\pi} d\phi r(\phi)  \nonumber \\
&\times \!\!\!
\int_0^\infty \!\!\! dE_2 \rho(E_2) \theta(E_1-E_2) 
\delta ( E_1' - F(E_1,E_2,\alpha, \zeta, \phi;c_r^*) )  , 
\label{eq:p}
\end{align}
where $Z(E_1)$ is fixed by the normalization condition, $1=\int dE_1'\rho(E_1' | E_1,c_r)$, and 
$F$ is 
\begin{align}
&
F(E_1,E_2,\alpha,\zeta,\phi;c_r^*)=E_{12} \big( \frac{1+c_r^2}{4} + \frac{1-c_r^2}{4} q \cos \zeta 
\nonumber \\
&+ \frac{c_r}{2}\sqrt{1-(q \cos \zeta)^2}   
\left(\cos \zeta \cos 2\alpha -\sin \zeta \sin 2\alpha \cos \phi \right)\big)\ ,
\end{align}
with $q = 2\sqrt{\frac{E_1}{E_{12}}\frac{E_2}{E_{12}}}$ and $c_r(\alpha)^2=1-(1-{c_r^*}^2)|\sin{\alpha}| $. 
The term $\theta(E_1-E_2)$ is introduced to account for the fact that the molecular chaos assumption is problematic for inelastic gases, and energy equipartition is generally not realized \cite{Feitosa2002} For details, see SI Text 3. Note that the transition probability functionally depends on the marginal energy distribution function. The expression for 2D transition probability is found in SI Text 4.

%\section{Results}

\begin{figure}[t]
\includegraphics[scale=0.15]{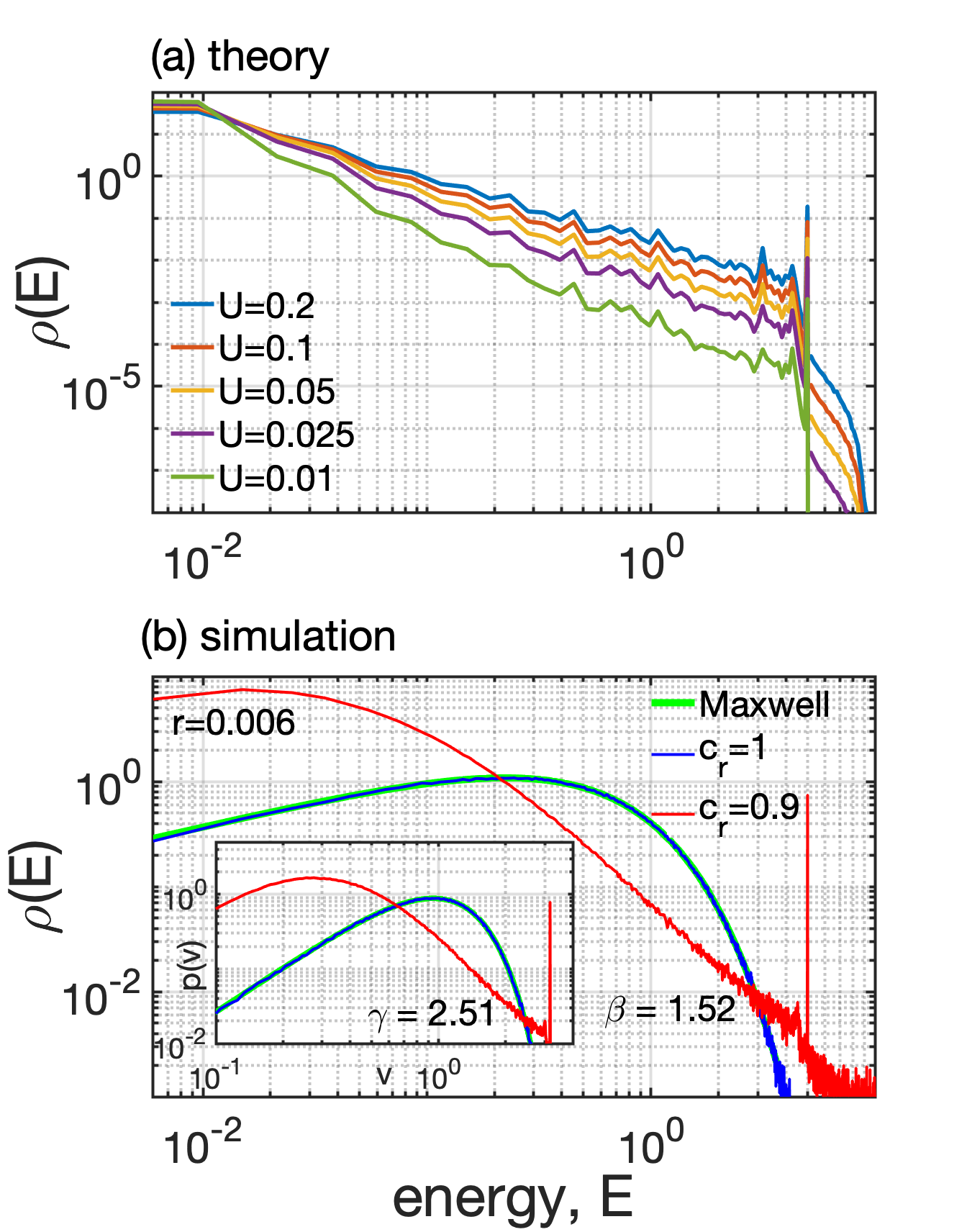}
\caption{Energy distribution of particles that collide inelastically,    
(a) as obtained from the numerical solution to Eq. (\ref{eq:e}) with Eq. (\ref{eq:p}) by fixing $c_r=0.9$ and various levels of internal energy, $U$. Clearly an approximate power law decay is visible, as well as the driving peak at $E=5$ that results from $\rho_{\rm charge}(E) = \delta(5)$. $U$ determines the driving rate, $r$.
(b) Energy distribution as obtained from a MD simulation of an ensemble of 125 particles. 
}
\label{fig:2}
\end{figure}
The self-consistent numerical solution to Eq. (\ref{eq:e}) with Eq. (\ref{eq:p}) is seen in Fig. \ref{fig:2} (a) for different values of the internal energy, $U$. For details, see SI Text 5. The choice of $U$ determines the driving rate, $r= \frac{2(\langle E\rangle_{\rm post}-U)}{\langle E\rangle_{\rm post}-\langle E\rangle_{\rm charge}}$. For the numerical solution, we fix $c_r$ and $U$. The charging energy distribution is set to a delta function,  $\rho_{\rm charge}(E) = \delta(E-5)$, i.e. particles receive a fixed energy, whenever charged. Clearly, the distribution is dominated by a power law, $\rho(E)\sim E^{-\beta}$, that extends over several decades of $E$. We fit the corresponding exponents, $\beta$, with a maximum likelihood estimator within appropriate bounds \cite{Hanel2017}. Also the driving peak at $E=5$ is visible. For energies above $5$, we see a much quicker drop in the energy distribution, a fact that has been described in \cite{Ben-Naim2005}. These high energy particles correspond to the relatively rare situation that a quick particle becomes faster in a collision. 

Figure \ref{fig:2} (b) shows the energy distribution, $\rho(E)$, of the system as obtained from a straight forward molecular dynamics (MD) simulation \cite{Haile1992} of $N=125$ particles with diameter $d=0.5$ in a 3D box of size $L=5$, and $c_r=0.9$. For making the driving compatible with the analytical computation, particles that hit a wall were reset to a constant energy of $5$ with a probability $\eta=0.5$, which resulted in an observed driving rate of $r=0.006$. For more details on the molecular dynamics simulation, see SI Text 6.  
Panel (b) shows a clear power law in the energy distribution (red), $\rho(E)$, very similar to panel (a). Also the driving peak and the steep fall-off for higher energies is visible. It is also visible that for low energies the energy distribution shows a deviation from the power law and forms a ``shoulder''. This is due to the geometric factors, that are of course also present for elastic collisions. The MD simulation for $c_r=1$ is show in blue and exactly follows the Maxwell-Boltzmann distribution (green), $\rho(E) = 2  \left( \frac{1}{k T} \right)^{\frac32} \sqrt{\frac{E}{\pi}} e^{- \frac{E}{kT}}$. The inset shows the velocity distribution, $c_r=0.9$ in red, $c_r=1$ in blue, green is 
$\rho(v) =   4 \pi \left( \frac{m}{2 \pi k T} \right)^{\frac32}  v^2   e^{- \frac{mv^2}{2kT}}$. The fact that the blue and green lines practically coincide demonstrates the quality of the MD simulation.  

\begin{figure}[t]
\includegraphics[scale=0.16]{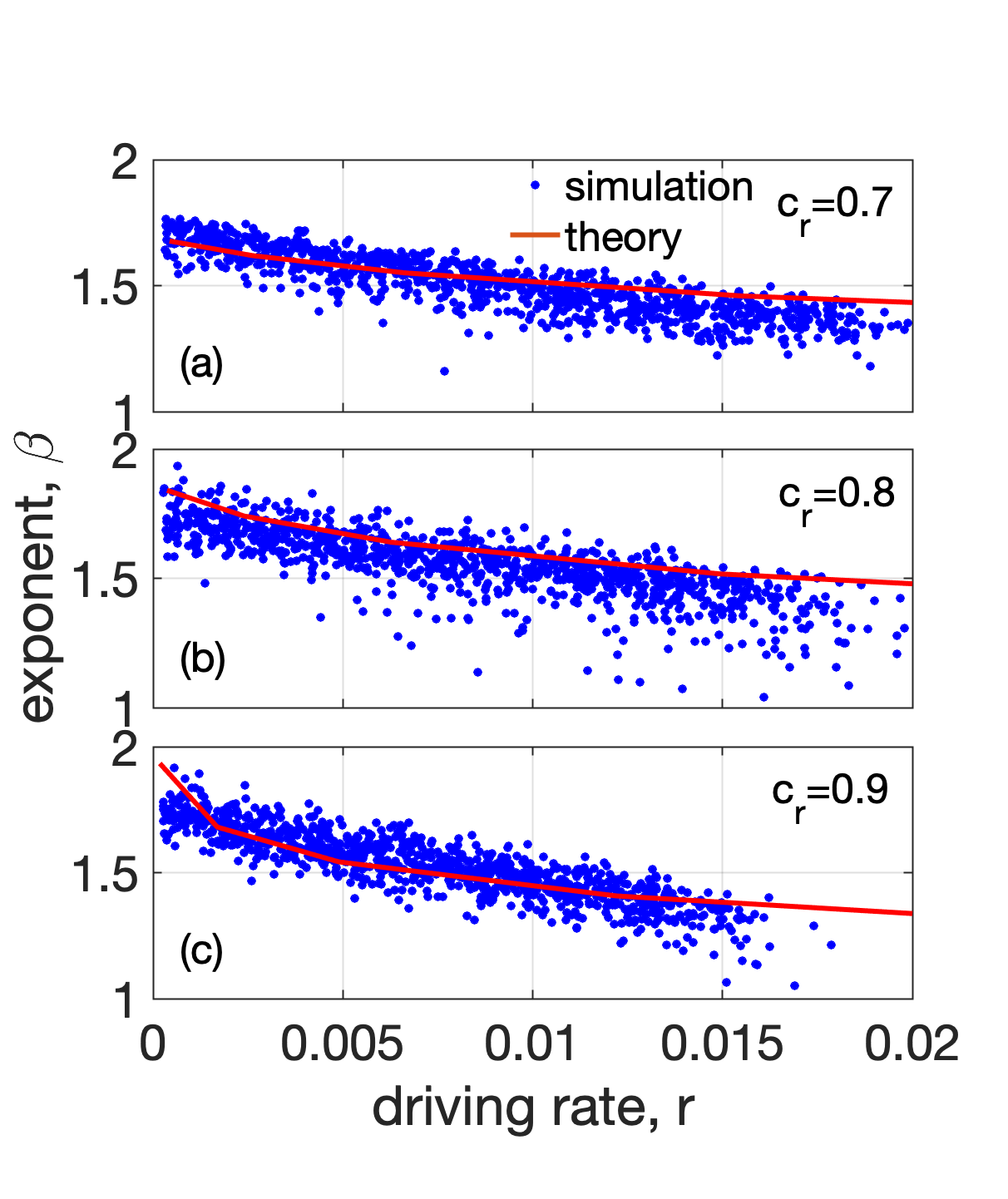}
\caption{ Dependence of the power exponent, $\beta$, on the driving rate, $r$, 
for (a)  $c_r=0.7$, 
(b)  $c_r=0.8$, and  
(c)  $c_r=0.9$.
Red lines indicate the SSR results, i.e., the solution to Eq. (\ref{eq:e}). Dots show the MD simulation. In both cases, the exponents were fitted with a least likelihood estimator \cite{Hanel2017} within appropriate fit-ranges that are specific to the different $c_r$. Note that also the lines have an error of about $\pm 0.01$ as a result of uncertainties in fitting.  
}
\label{fig:3}
\end{figure}

In Fig. \ref{fig:3} we show the dependence of the exponent, $\beta$, that solves Eq. (\ref{eq:e}) as a function of the driving rate, $r$ (red line), see SI Text 5. Clearly, $\beta$ is below $2$, and decreases with increasing driving. The situation is shown for $c_r=0.7$, $0.8$, and $0.9$. The larger $c_r$, the steeper exponents decline. Note, that exponents are fitted to distributions like the one shown in Fig. \ref{fig:2} (a) and thus contain an error of $\pm 0.01$ that is due to the fitting procedure \cite{Hanel2017}. The blue dots are the results from the MD simulations, that were realized by varying $\eta$ from $0.02$ to $1$ in steps of $0.01$. For each condition, 10 independent runs of $200,000$ collisions were performed before fitting $\beta$. The spread in the simulation shows the variability and errors in the estimation of $\beta$. In every individual run, the driving rate was determined as the actual number of charging events per actual particle-particle collision. Generally, for 2D we find qualitatively very similar results. 
\begin{figure}[t]
\includegraphics[scale=0.16]{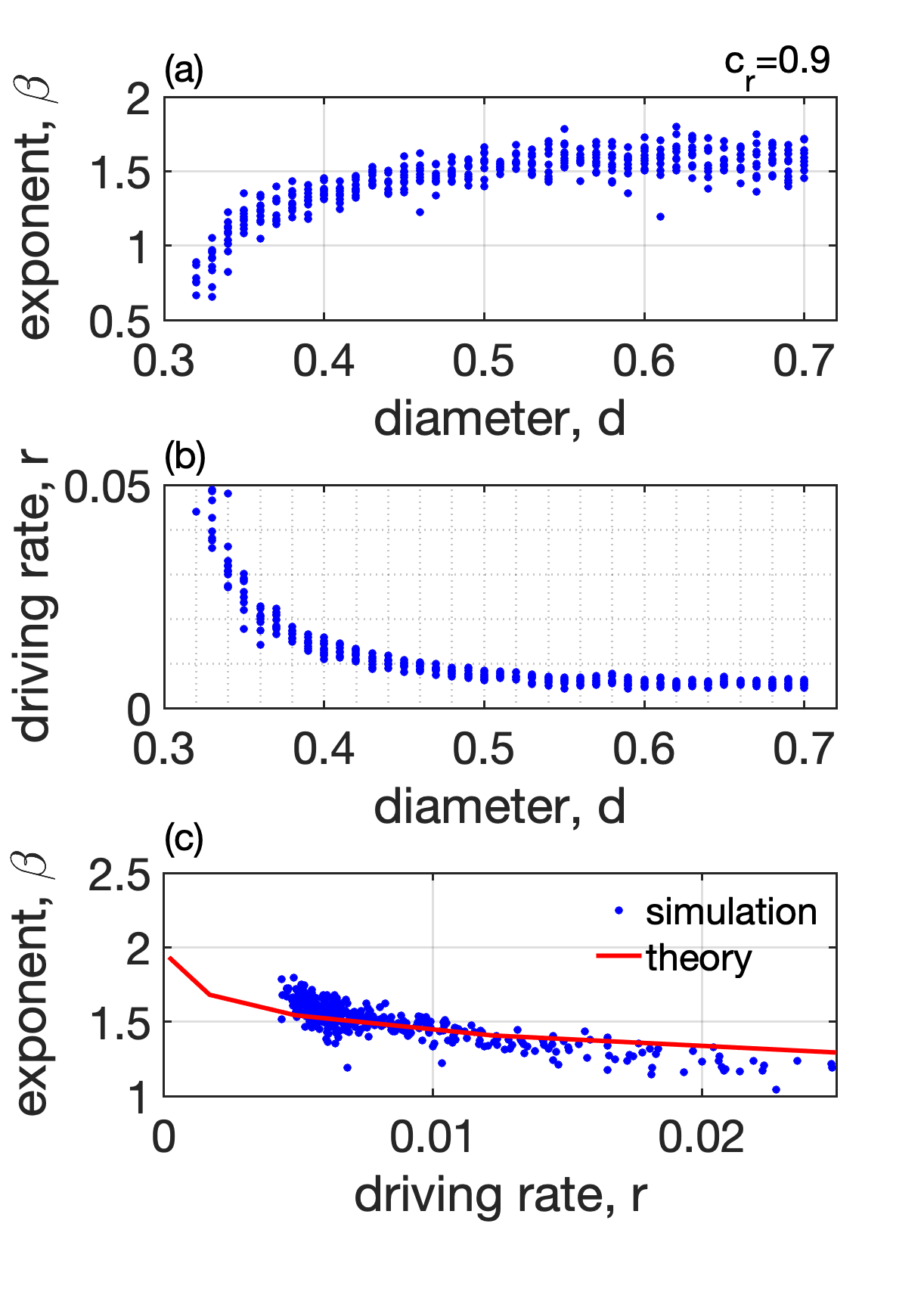}
\caption{Dependence of the exponent on geometrical factors. 
(a) $\beta$, increases as a function of the diameter, $d$, however, the driving rate decreases as seen in (b), since particle-particle interactions become more frequent. Symbols show MD simulation results for $c_r=0.9$.
(c) When the exponent is plotted against the respective driving rate we obtain the previous SSR result (red line). This confirms that the effect of particle size translates to the frequency of particle-particle collisions and the driving rate can be identfied as the relevant parameter.
}
\label{fig:4}
\end{figure}
We note a dependence of the exponents on geometrical parameters, such as the diameter, $d$, as we show in Fig. \ref{fig:3} (a).
Larger particles collide more often, thus the driving rate, $r$, decreases with increasing $d$,  see Fig. \ref{fig:3} (b). If one plots the exponent, $\beta$, versus $r$, the theoretical result (red line) is holds, see Fig. \ref{fig:3} (c). 
If we assume the existence of pure power laws between the ``shoulder'' and the driving energy, the relation $\gamma=2\beta-1$
between the velocity and energy $\beta$ exponents should hold, see SI Text 7. In SI Fig. 4 we show that it does.

%\section{Discussion}

We generalized the Maxwell-Boltzmann distribution for inelastic gases in their simplest form. We  demonstrated that it is possible to derive the energy distribution for inelastic gases in the framework of sample space reducing processes that are characterized by a specific type of eigenvalue equations  that have been associated with universal power laws if systems are driven slowly and in a simple manner. We derived the corresponding equation that incorporates the collision geometry and a simple driving process, where particles are energized whenever they hit the walls of the box. 

We could demonstrate that the bulk of the energy (and velocity) distribution function, follows a clear power law that depends on the coefficient of restitution and the driving rate of the system. We confirm these findings with a straight forward molecular dynamics simulation that leads to practically same results; we showed them in 3D and checked that they remain valid also in 2D. 

We emphasize that this is a very different approach to inelastic gases than the route through the (non-linear) Boltzmann equation, as e.g. in \cite{Ben-Naim2005}. Also there, scaling distributions were obtained, however, with significantly larger exponents 
of around $-5$ for $c_r = 0$ and $-6$ for $c_r \sim 1$ in 3D, that are valid for the super fast particles only and do not characterize the entire distribution. We can confirm, however, for particles above the driving energy that the energy distribution is compatible with the steep decrease found in \cite{Ben-Naim2005}. The assumption of a weak molecular chaos used in \cite{Ben-Naim2005} is valid only for the very high energy particles. It is not valid, however, for lower energies that describe the vast majority of the particles in the system. The use of the Boltzmann equation is problematic (even its non-linear version) since post-collision velocities can't be considered independent. Instead, a large system of BBGKY hierarchy equations should be used \cite{Spohn}, for which moment closure obtained from the Boltzmann-Grad limit does not lead to a single Boltzmann equation. The presented alternative of the SSR approach overcomes the difficulty of solving the system of BBGKY hierarchy equations and offers a solution for all velocity scales. The present result is a clear demonstration the the SSR framework is powerful enough to compute  energy and velocity distribution functions for concrete driven dissipative systems, at least for relatively simple ones. 

\begin{acknowledgments}
This work was supported in part by Austrian Science Fund FWF under P29032 and P33751, 
and the Austrian Science Promotion Agency, FFG project under  873927. 
\end{acknowledgments}

\clearpage
% \appendix
 
 \onecolumngrid

 \section*{Supplemental Material}

\subsection*{SI Text 1: Derivation of the single-particle transition function} 

%%%% TO SI
%%%%%
In the laboratory frame, the energies of two colliding particles are $E_1$ and $E_2$. We follow a reference particle 1 through the collision with energy $E_1$ before and $E_1'$ after the collision. To see what happens to the particle we consider particle 2 with energy $E_2$ drawn from the energy distribution $\rho(E)$ of the gas. We now compute the transition probability $\rho(E_1'| E_1,c_r)$. 

$c_r$ may depend on the reflection angle, $2\alpha$. $c_r$ is the ratio of the velocity before and after a collision in the center of mass system, thus $c_r^2$ corresponds to the ratio of energies before and after a collision. In Fig. \ref{fig:1}(a)  it is easy to see that the tangent angle to the velocity vector, $\alpha$, for the two colliding particles is $\cos(\alpha)={h}/{d}$, where $h$ is the displacement between the particle centres orthogonal to the relative velocity. We assume that for inelastic collisions the reflection angle is $2\alpha$, as for the elastic case. However, there exists a monotonic function $c_r(\alpha)$ for $\alpha\in \left[0,\pi\right]$, that has a minimum for $c_r^*\equiv c_r (\pi/2)$ and $1=c_r(0)=c_r(\pi)$. $\nu(\alpha)=1-c_r(\alpha)^2$ plays the role of an energy dissipation factor; it is a monotonic increasing function. A natural Ansatz is $\nu(\alpha)=(1-{c_r^*}^2)|\sin{\alpha}|$ and $c_r(\alpha)^2=1-(1-{c_r^*}^2)|\sin{\alpha}| $. 

To compute the pair-energy distribution before and after a collision for the angle dependent $c_r(\alpha)$, one can proceed as follows. The energy of the particle-pair is given by $E_{12}=\frac12 m\left(v_1^2+v_2^2\right)$. In center of mass frame after the collision we have $ {{E_{12}}^{\rm cm}}'  ={c_r}^2  {E_{12}}^{\rm cm}$. In the laboratory frame this is $E_{12}'=\frac{1+c_r^2}{2}E_{12}+\frac{1-c_r^2}{2}m(v_1|v_2)$, where the scalar product is $(v_1|v_2)=|v_1||v_2|\cos \zeta =\frac{2}{m}\sqrt{E_1E_2}\cos \zeta$. $\zeta\in[0,\pi]$ is the angle between the velocity vectors. Since $E_{12}=E_1+E_2$, $E_1E_2$ takes a maximum at $E_1=E_2=E_{12}/2$, so that the maximal value of $\sqrt{E_1E_2}$ is $E_{12}/2$ and it follows that $m|(v_1|v_2)|\leq E_{12}$. As a consequence, the range the values of $E_{12}'$ are restricted to the interval $[c_r^2 E_{12},E_{12}]$, i.e. $c_r^2 E_{12}\leq E_{12}' \leq E_{12}$. Since the most likely value of $E_1E_2$ is also where $\sqrt{E_1E_2}$ is maximal, we use the Ansatz $m(v_1|v_2)=qE\cos \zeta$, with $q(E_1,E_2)\equiv 2\sqrt{\frac{E_1}{E_{12}}\frac{E_2}{E_{12}}}$. $0\leq q\leq 1$, with the most likely value being at $q=1$. We finally get $E_{12}'= \left ( \frac{1+c_r^2}{2}+\frac{1-c_r^2}{2}q\cos \zeta \right) E_{12}$, where the dependence on the particular initial kinetic energies is absorbed into the random variable $q$.

To compute the energy of one particle after the collision in 2D, we express it in terms of the prior energies, $E_1$ and $E_2$, the angle between velocities prior to the collision, $\zeta$, and the reflection angle, $2\alpha$. Since we know how the total energy $E_{12}$ behaves, we just have to calculate $E_1'$ after the collision. For this we rotate the laboratory coordinates such that  $v_1$ and $v_2$ are in the $xy$-plane and the center of mass velocity $v_{\rm cm}=(v_1-v_2)/2$ of particle $1$  points in $x$ direction. The velocity of the mass center, $u=(v_1+v_2)/2$, has an angle $\zeta$ with $v_{\rm cm}$. For a picture of the geometry, see SI Fig. 1(a). We write
\begin{equation}
|v_1'|=\left|  |u|\, \left(
\begin{array}{c} \cos \zeta \\ -\sin \zeta \end{array} 
\right)\ +\  c_r \, |v_{\rm cm}|\,  \left( 
\begin{array}{c} \cos \, 2\alpha \\ \sin \, 2\alpha \end{array} 
\right)\, \right|\ .
\end{equation}
Using this Ansatz it follows that
\begin{equation}
|v_1'|^2=|u|^2+c_r^2|v_{\rm cm}|^2+2c_r |u||v_{\rm cm}|\left(\cos \zeta \cos 2\alpha -\sin \zeta \sin 2\alpha \right)
\end{equation}
and one arrives at 
\begin{eqnarray}
\frac{E_1'}{E_{12}} &=&\frac{1+c_r^2}{4}+\frac{1-c_r^2}{4}q\cos \zeta 
+\frac{c_r}{2}\sqrt{1-(q\cos \zeta)^2}    
%\nonumber \\ 
%&\times&  
\left(\cos \zeta \cos 2\alpha -\sin \zeta \sin 2\alpha \right)\, .
\label{eq:singleparticleetrans2d}
\end{eqnarray}
For 3D we introduce a rotation angle $\phi\in[0,\, \pi]$ of the center of mass velocity of particle 1 after the collision in the $yz$-plane, see SI Fig. 1 (b), and get 
\begin{eqnarray}
\frac{E_1'}{E}&=&\frac{1+c_r^2}{4}+\frac{1-c_r^2}{4}q\cos \zeta
+\frac{c_r}{2}\sqrt{1-(q \cos \zeta)^2}   
% \nonumber \\ 
%&\times& 
\left(\cos \zeta \cos \, 2\alpha- \sin \zeta \sin 2\alpha \cos \phi \right) \, .
\label{eq:singleparticleetrans3d}
\end{eqnarray}

Next, we compute the distribution function of the reflection angle, $2\alpha$, assuming isotropic conditions. We need the probability of two colliding particles to be at an orthogonal distance, $h$. In 2D, all $h\in \left[0,D\right]$ are equally likely, since there are just two possibilities ($h$ and $-h$), and $\rho(h)=1/d$. In 3D, $\rho(h)=2h/d^2$, since the area of collisions with orthogonal distance, $h$, is $dA=2\pi\, h\, dh$. To get the probability distribution for angle $\alpha$, $\rho(\alpha)$, one requires $|\rho(\alpha) d\alpha|=|\rho(h) dh|$, and it follows that for 2D, $f(\alpha)=|\sin \alpha |$ and for 3D, $f(\alpha)=|\sin 2\alpha |$. In 2D, half of the colliding particles are reflected with $2\alpha$, the other half is reflected at $-2\alpha$. In 3D, the collision plane can be rotated between $0$ and $2\pi$. Since the mirror reflection, $2\alpha\ \to\ -2\alpha$, is not equivalent to the reflection with angle $2\alpha$, we extend the domain from $\alpha\in[0,\pi/2]$ to $\alpha\in[0,\pi]$, with a respective renormalisation of the probabilities in $f(\alpha)$. 
To compute the distribution of angles between $v_1$ and $v_2$, under the assumption of isotropy, every angle, $\zeta$, has only one possibility to be realized and the distribution is uniform, $g(\zeta)=1/\pi$. In 3D, fixing $v_1$, there is a rotational degree of freedom for $v_2$ that has a fixed angle with $v_1$, and $g(\zeta)=\sin(\zeta)/2$. Finally, for the distribution of $\phi$, that only exists in 3D, we safely assume it to be uniform in $[0,\pi]$, and $r(\phi)={1}/{\pi}$.
%%%%% 
%%%%% TO SI 
%%%%% 

\subsection*{SI Text 2: Geometry of the collision} 
For the computation of $E_1$ in terms of the initial velocities, $v_1$ and $v_2$, the collision angle, $2\alpha$, and the angle $\zeta$.  
We depict the situation for 2 dimensions in 
SI Fig. 1 (a). the case for 3 dimensions is in panel (b).
\begin{figure}[ht]
\begin{center}
		(2d) \includegraphics[width=0.3\columnwidth]{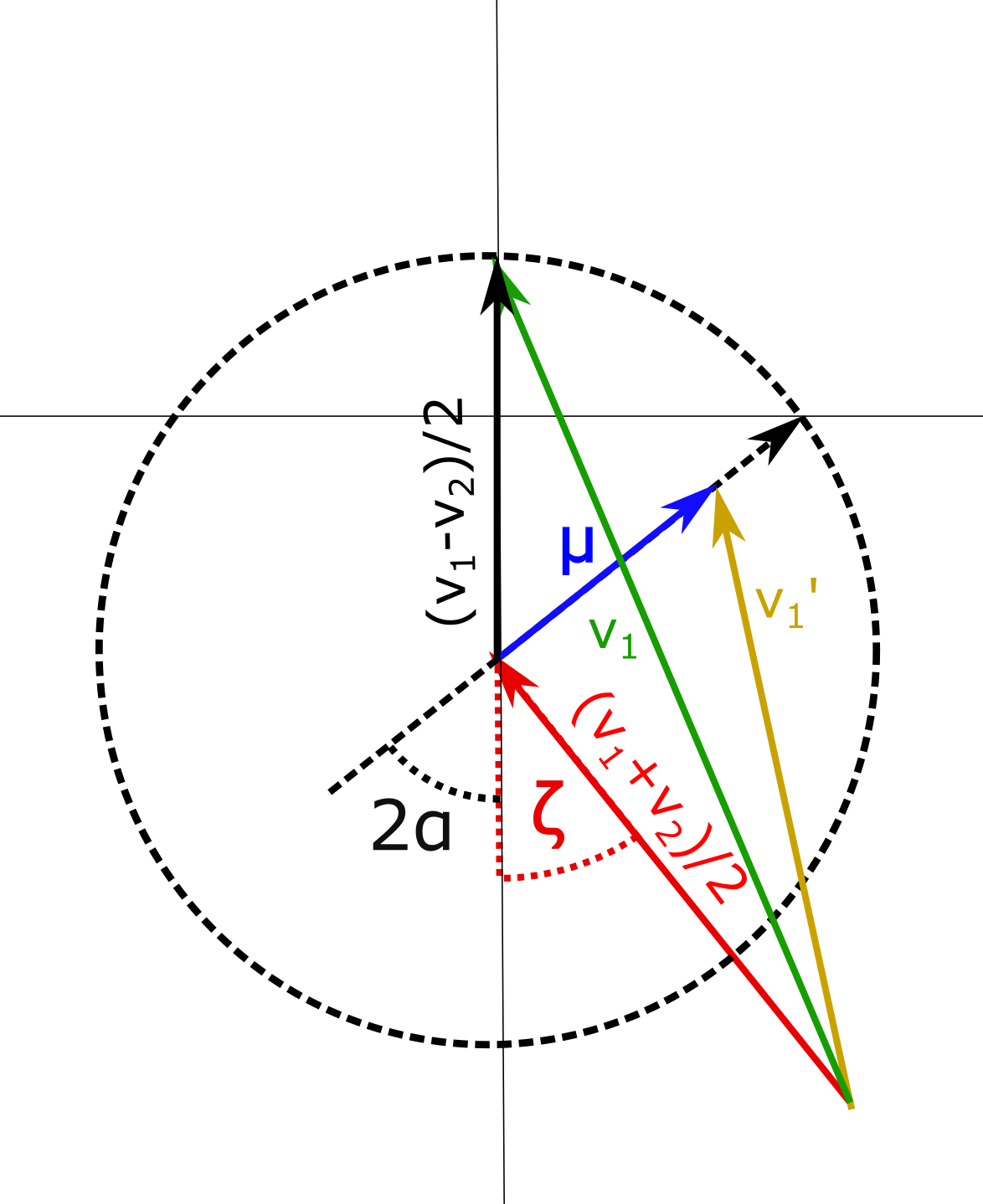}
		(3d) \includegraphics[width=0.3\columnwidth]{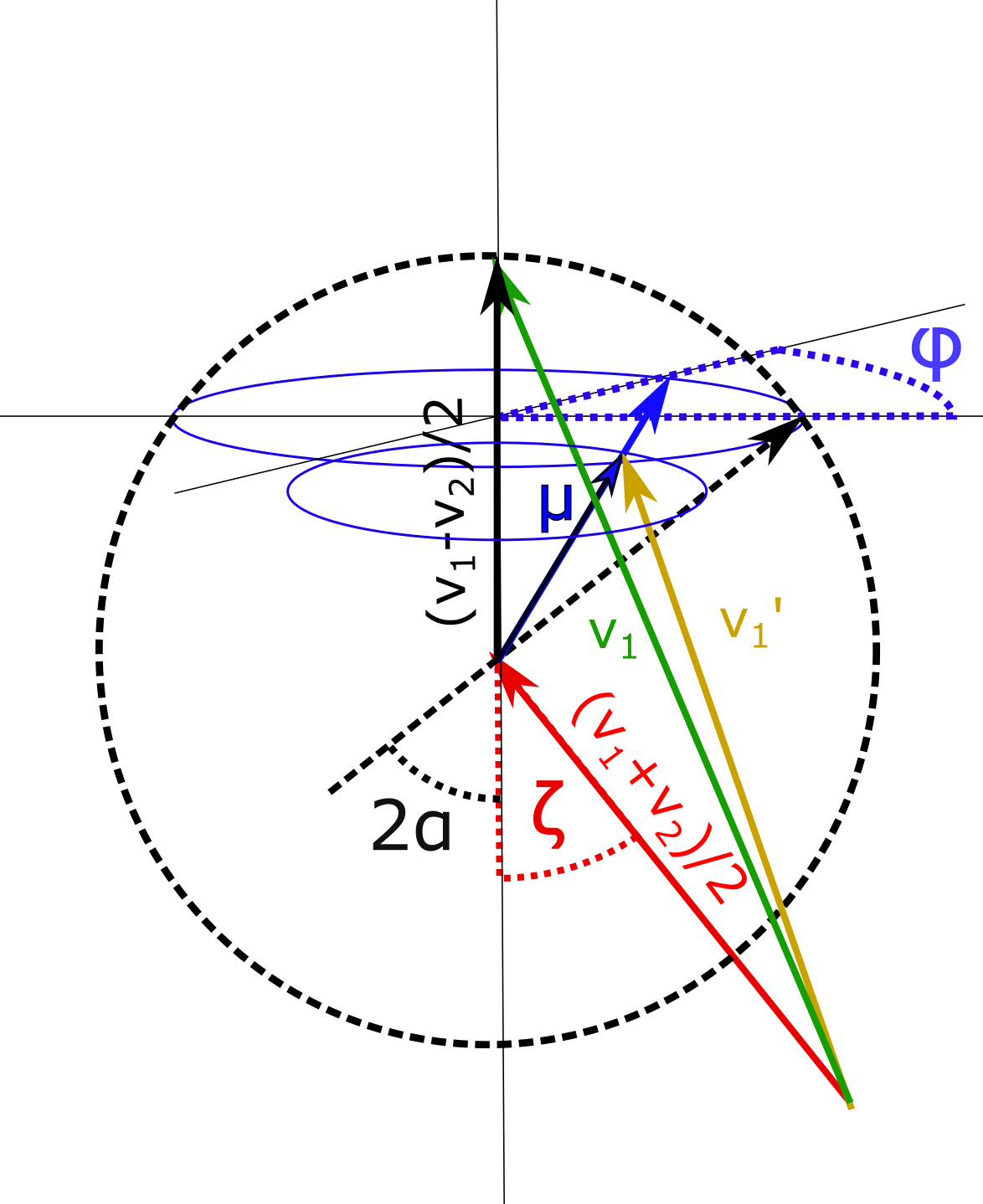}
\end{center}
	\caption{{\bf SI Fig. 1} 
	(a) Collision geometry in terms ov velocities and angles for 2D.
	(b) Case for 3D with the additional angle $\phi$.
}	 	
	\label{fig:collision2d3d}
\end{figure}

\subsection*{SI Text 3: Note on the molecular chaos assumption in dissipative systems} 

The molecular chaos assumption (that velocities of particles are well mixed) does not hold particularly well in inelastic collisions. 
It is well known that systems working irreversibly between an energy source and an energy sink can decrease their entropy, which in the simplest form is due to  deviations from the uniform distribution of micro-states (equipartition property), caused by the energy current in the system. It is also known that non-equilibrium steady states have cycles (cycling theorems) and break local detailed balance, i.e., we can't assume that the probability of observing two particles with velocities $v_1$ and $v_2$ together in a collision, factorizes into the marginal probabilities of observing $v_1$ and $v_2$ separately. In fact, if a particle receives an energy boost at the wall through the charging process, it has high energy (high velocity), and will quickly collide with a slower particle. Other slow particles can no-longer collide with this once high energy particle, since it already dissipated the energy to particles with typically lower energies. 
To assume the same free path-length --or alternatively, the same inter particle collision times-- for all particles is thus unrealistic. Slow particles will practically freeze out into clusters of slow particles, as observed in \cite{Kudrolli1997}. From time to time a high energy particle will hit such a cluster, and dissipate its energy to the cluster.

\subsection*{SI Text 4: Formula for the 2D single-particle transition probability} 

\begin{eqnarray}
\rho(E_1'|E_1,c_r) &=&\ \frac{1}{Z(E_1)}\int_{0}^\pi d\zeta\ g(\zeta) 
\int_0^{\pi} d\alpha\ f(\alpha) \int_0^\infty dE_2\ \rho(E_2)
\delta\left( E_1' - F(E_1,E_2,\alpha,\zeta,0;c_r^*) \right) \\
\label{eq:slowandfast2D}
\end{eqnarray}

\subsection*{SI Text 5: Details on the numerical solution of the eigenvalue equation} 

To solve the equation numerically in reasonable time one must introduce a high energy cut-off and discretize the domains of the various integration variables in a relatively coarse way. 
For the solution, we chose to fix the internal energy $U$, that  
\begin{equation}
U=(1-\xi) \langle E\rangle_{\rm post}\ +\  \xi \langle E\rangle_{\rm charge} \, ,
\label{eq:U}
\end{equation}
where $\langle E\rangle_{\rm post}=\int_0^{\infty} dE E\ \rho_{\rm post}(E)$, 
with $\rho_{\rm post}(E)=\int_{E_0}^{E^*}dE\rho(E'|E,\mu)\rho(E)$, 
(all appropriately discretized).
For the charging process we chose to re-introduce particles at a fixed energy, i.e. we set the energy distribution for the 
charging process to a Dirac-delta at $E=5$, and get $\langle E\rangle_{\rm charge} =5$.
Since $U$ is fixed and $\langle E\rangle_{\rm post}$ is computed in the algorithm, by using Eq. (\ref{eq:U}) 
we get the driving rate, $r=2\xi=\frac{2(\langle E\rangle_{\rm post}-U)}{\langle E\rangle_{\rm post}-\langle E\rangle_{\rm charge}}$. 

The eigenvalue problem was performed in the following way. We appropriately discretized the integral domains of the angles. We bin the domains of the respective angles $\alpha$, $\zeta$, and $\phi$ into equal sized domains and use the bin-centers as the discrete angle values used in the sums approximating the integrals over the respective angles in the energy eigen-distribution equation.  For $\alpha$ we use 13 bins, for $\zeta$ and $\phi$ 9 bins each. Using odd numbers of bins avoids the necessity of dealing with expressions of the form $0/0$ in the formulas and the need for analysing and implementing the defined limits  corresponding to the situation $x/y\to 0/0$ .The energy domain is more involved. First, we have to keep the number of bins low in order to respect constrains of computing time. At the same time, we would like to allow the internal energy, $U$, to be small, which implies that the bin-size for low energies needs to be small, and the high energy cut off, $E_{\rm max}=50$ (remember $E_{\rm charge}=5$). We can accomodate all three criteria by using not equally spaced energy bins for the energy domain. We use $N=300$ bins and place them in the following way
\begin{equation}
\epsilon_n=\gamma \left((a^2+n^2)^{1/2}-a\right) \, ,
\end{equation}
with $a=40$ and $\gamma$ chosen such that $\epsilon_N=E_{\rm max}$.

The eigenfunction equation then is solved iteratively initialising the particle energy distribution function, $\rho(E)$, uniformly distributed on the energy interval $\left[0,2U\right]$, and vanishing outside of it. The procedure converges quickly and we use a fixed number of 7 iterations for obtaining our results, which we checked, is sufficient for our purposes. In each iteration we compute the energy transition distribution once. However, we iteratively update the energy distribution $\rho(E)$ three times using the same energy transition probability so that we effectively iterate $\rho(E)$ for $21=3\cdot 7$ times for the solutions we obtain.

For computing the eigenfunction problem, we choose an energy threshold $E_{T}=20$ and consider only energy bins below that threshold for the eigenfunction equation of the discretised distribution function $\rho(\epsilon_n)$. The energy bins $\epsilon_n$ between $E_{T}$ and $E_{\rm max}$ are only used for estimating weight located in the tail of the distribution in order to minimise the deviation of energy expectation values induced by the energy cut off at $E_{T}$, i.e. we approximate $\rho(E?|E)$ as a rectangular transition matrix $\rho(\epsilon_m|\epsilon_n)$ with $0<\epsilon_m<E_{\rm max}$ and  $0<\epsilon_n<E_{T}$. However, only the part 
$0<\epsilon_m<E_{T}$ and  $0<\epsilon_n<E_{T}$ is used for solving the eigenfunction problem.
The remaining part of the matrix, $E_{T}<\epsilon_m<E_{\rm max}$ and  $0<\epsilon_n<E_{T}$ is collected only for roughly estimating the tail of the energy distribution function, $\rho(\epsilon_m)$. The respective MATLAB codes are made available.

\subsection*{SI Text 6: Computer simulation of inelastic particles in a box} 

We use a standard molecular dynamics (MD) scheme for spherical particles with diameter $d$ in finite box of length $L$ in 1, 2, and 3 spatial dimensions \cite{Haile1992}. For simplicity, we set all masses equal, $m=1$. Particles are initialized at random positions in the box with velocities in random directions. The absolute value of the initial velocity is taken from a uniformly distributed random number between 0 and 2. We distinguish between particle-particle collisions and particle-wall collisions. The update happens collision by collision. We compute the next particle-particle or particle-wall collision. For a particle-particle collision we update the velocities according to Eq. (2) in the main text, taking the coefficient of restitution into account. For a particle-wall collision particles are reflected off the wall as if it were an elastic collision, i.e. the directions after the wall collision are as for elastic reflections. For the base scenario with probability $\eta$ we choose the particle for an energy update and set it to a fixed kinetic energy, $E_{\rm charge}$. After every update  (particle-particle or particle-wall) the next collision is computed. The system typically converges to a reasonably steady (energy) state after a few thousand collisions, see SI Fig. 2. For the simulations we typically compute a few million collisions after removing the first 10.000 collisions.   

  \begin{figure}[ht]
\includegraphics[scale=0.33]{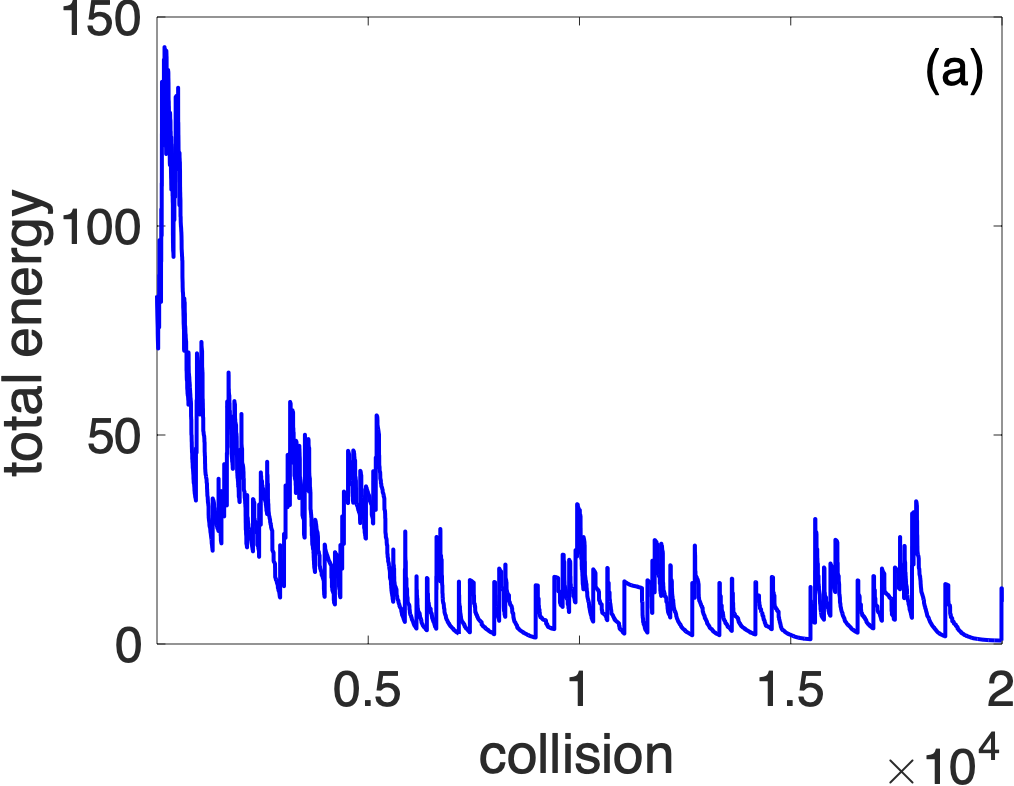}
\includegraphics[scale=0.33]{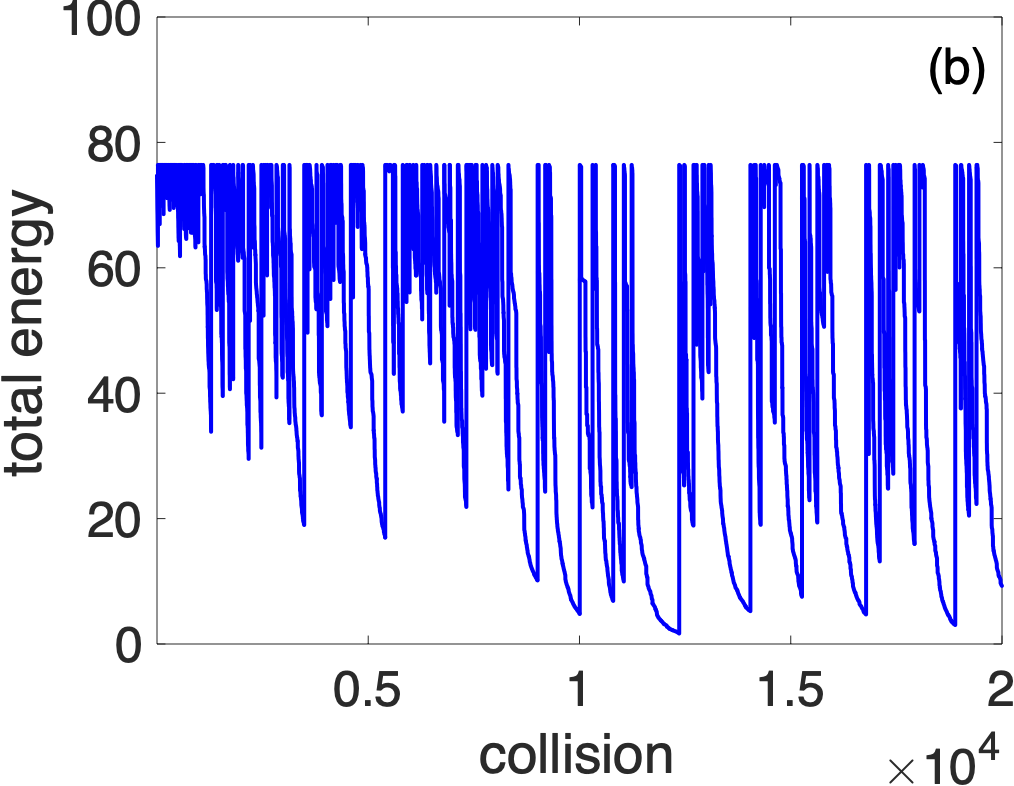}
\caption{{\bf SI Fig. 2} 
(a) Total energy in the system as a function of collisions. Every 'time step' corresponds to one collision (particle-particle, or particle-wall collision). Wall collisions drive the energy up by a fixed amount, particle-particle collisions dissipate the energy. Steady state is reached after about 5000 collisions.
(b) Situation for the  alternative energy update, where the dissipated energy is re-introduced to particles such that the system gets back to its initial energy after every driving event.
}
\label{fig:SM2}
\end{figure}

We implemented  an alternative energy update where energy is re-charged at wall hits with probability $\eta$, however, with exactly that energy that was lost in all the particle-particle collisions since the last charging process. In this way, the system is pushed back to its initial energy level after every recharging event. We found little effect in the distribution functions when using this alternative, see SI Fig. 3. 

\begin{figure}[ht]
\includegraphics[scale=0.15]{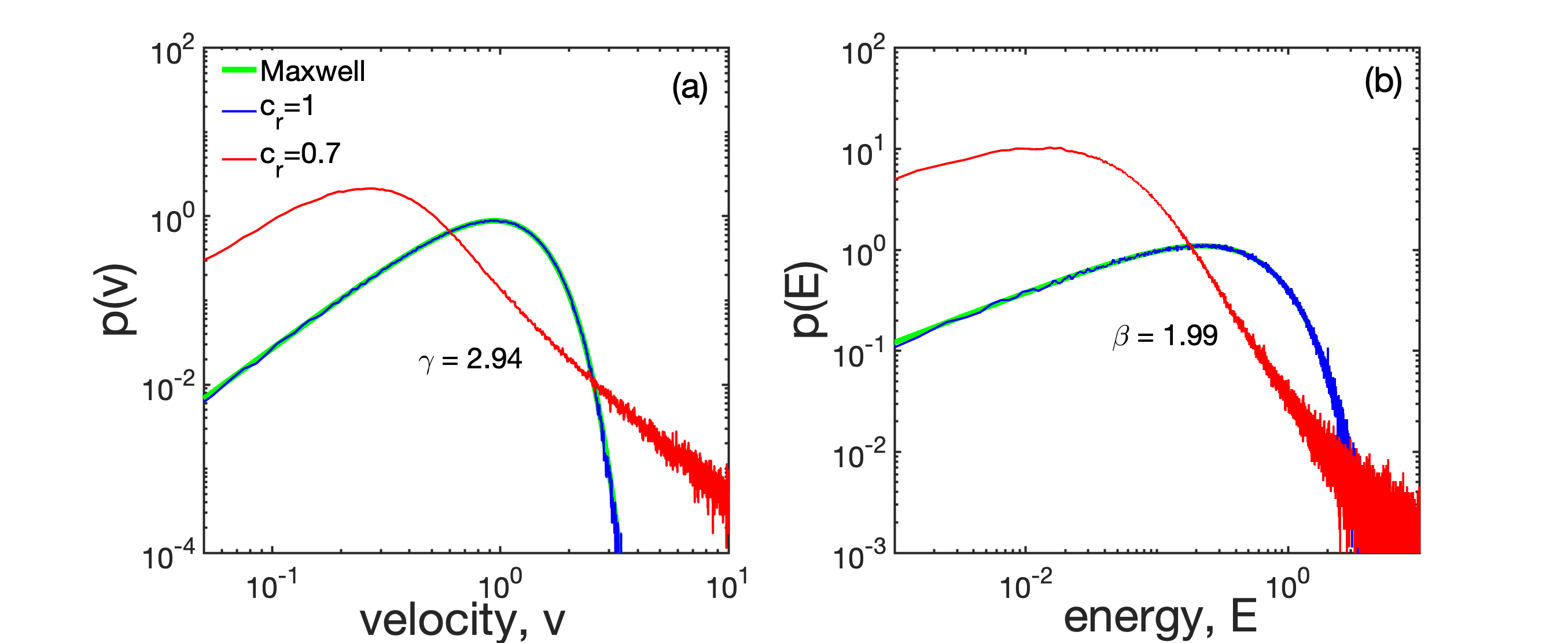}
\caption{{\bf SI Fig. 3} 
Distribution functions for the alternative driving scheme (red) still show extended power laws for 
(a) the velocity distribution, and  
(b) the energy distribution.
Compare with  Fig. 2 in the main text. Blue curves is MD  simulation for elastic collisions, $c_r=1$, green is the exact Maxwell-Boltzmann result. 
$c_r=0.7$. 
}
\label{fig:SM3}
\end{figure}

 \subsection*{SI Text 7: Convolution product of power-laws -- relation between $\beta$ and $\gamma$} 

The convolution product, $p^*$, of power laws of type, $p(x)\propto x^{-\alpha}$, is written as 
\begin{equation}
p^*(x)=\int_{x_0}^x dx'\ p(x')p(x-x')=x^{1-2\alpha}\int_{x_0/x}^1 dz\ f(z)f(1-z) \, , 
\end{equation}
where the last factor does not depend on $x$. Also a numerical analysis shows that the tail of the distribution follows a power law with the expected exponent, $p^*(x)\propto x^{1-2\alpha}$ very nicely if $x_0$ is small ($x_0\sim 0.1$). In other words, if $x_0$ is small, $p^*(x)\propto x^{-\beta}$, with $\beta=2\alpha-1$.

This means the following for the energy and velocity distributions, $\rho(E)$ and $p(v)$.  Assume that $q(\mathbf{v})=f(|\mathbf{v}|)$ is an isotropic particle velocity distribution with $\mathbf{v}\in \mathbb{R}^3$. We use  $v=|\mathbf{v}|$ for the absolute value of the velocity vector; $E=m v^2/2$ is the kinetic energy of the particles.
\begin{equation}
\begin{array}{lcl}
1&=&\int_{|u|\leq v} d^3\mathbf{v}\ q(\mathbf{v})\\
&&\\
&=& 4\pi\int_0^v dv'\ |v'|^2f(v')\\
&&\\
&\equiv &\int_0^v dv'\ p(v')\\
&&\\
&=&\int_0^v d(\sqrt{2E'/m})\ p(\sqrt{2E'/m})\\
&&\\
&=&4\pi\int_0^{mv^2/2} \sqrt{\frac{2}{m}}\frac{dE'}{2\sqrt{E'}}\ \frac{2E'}{m}\ f(\sqrt{2E'/m})\\
&&\\
&=&2\pi\left(\frac{2}{m}\right)^{3/2}\int_0^{mv^2/2} dE'\ \sqrt{E'}f(\sqrt{2E'/m})\\
&&\\
&\equiv &\int_0^{E} dE'\ \rho(E')\ .
\end{array}
\end{equation}
By comparing terms in the transformation of variables we can identify
\begin{equation}
\begin{array}{lcl}
\rho(E) &=& 2\pi\left(\frac{2}{m}\right)^{3/2}\sqrt{E}f(\sqrt{2E/m})\\
&&\\
 p(v) &=& m|v| \rho(m v^2/2)\ .
\end{array}
\end{equation}
If we observe a power law $\rho(E)\propto E^{-\beta}$, then we necessarily observe a power law $p(v)\propto v^{-\gamma}$ with  $\gamma=2\beta-1$. Also $f(v)\propto v^{-(\gamma+2)}\propto v^{-(2\beta+1)}$ holds. 

Note that the relation between  exponents $\beta$ and $\gamma$ is identical to the relation between a power law distribution (e.g. the one particle energy distribution function $\rho(E)$) and its convolution product with itself, the distribution function of two particles $\rho^*(E)$.
That is, if $p(v)\propto v^{-(2\beta-1)}$ it has the same exponent as 
\begin{equation}
\rho^*(E)=\int_0^{E} dE' \rho(E-E')\rho(E') \propto E^{-(2\beta-1)}\ . 
\end{equation}

In SI Fig. 4 we see the realization of the expected relations between the two exponents, $\beta$ from the energy and $\gamma$ from the velocity distribution. The expected relation $\gamma=2\beta-1$ is realized approximately in the numerical MD simulations. 

\begin{figure}[tb]
\includegraphics[scale=0.15]{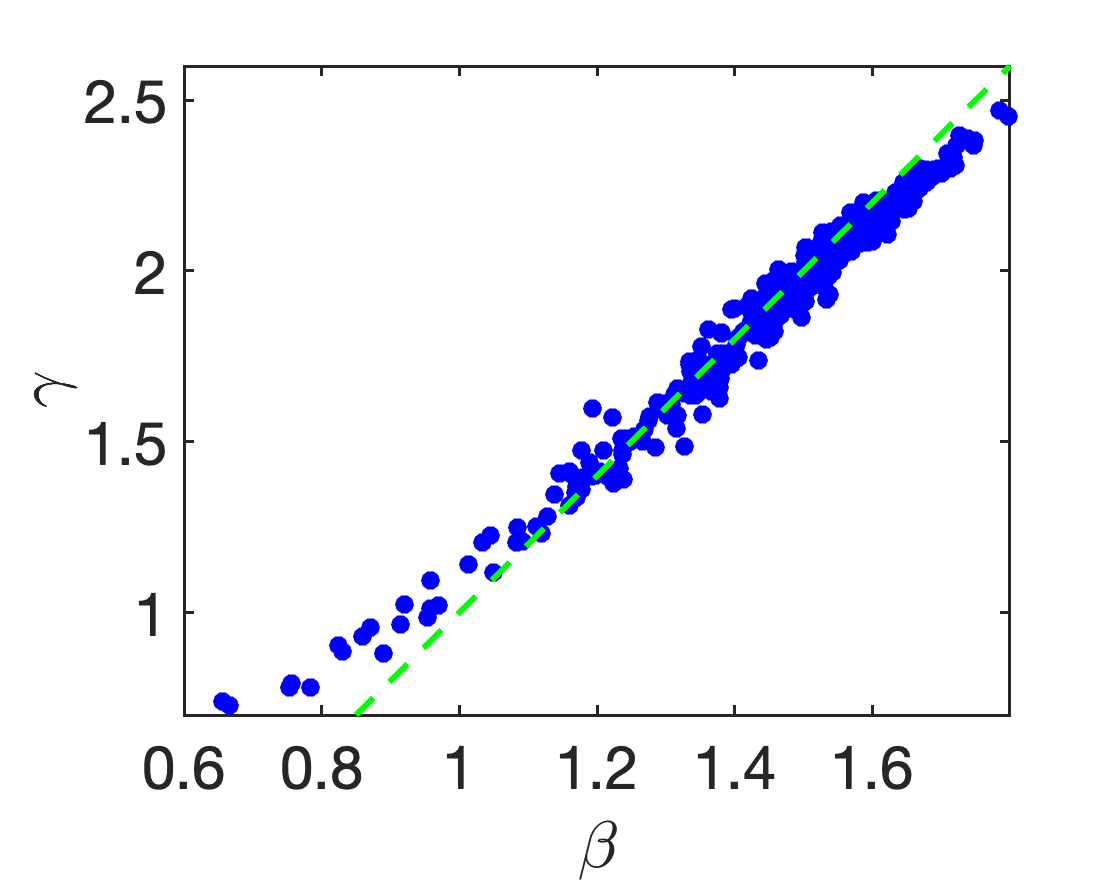}
\caption{ {\bf SI Fig. 4}
$\beta$  versus $\gamma$ exponent , 
for the numerical MD runs shown in Fig. 4 in the main text.  $r=0.9$. The dashed green line indicates 
$\gamma=2\beta-1$. The deviations are due to the fact that we employ fixed fit-regions, which are not 
equally optimal for all values of $d$.  
}
\label{fig:4b}
\end{figure}

\end{document}